\documentclass[11pt,a4paper]{article}
\usepackage{amsmath,amssymb}
\usepackage{color}
\usepackage{graphicx}
\usepackage{subfig}
\usepackage{cite}                
\usepackage{hyperref}            
\usepackage{multirow,makecell}
\usepackage{braket,bm}
\usepackage{cancel}
\RequirePackage{doi}
\usepackage{hyperref}
\usepackage{leftidx}
\usepackage[text={17cm,24.5cm},centering]{geometry}
\usepackage{textcomp}
\usepackage{wasysym}
\usepackage{mathtools}
\numberwithin{equation}{section}

\def \be {\begin{equation}}
\def \ee {\end{equation}}
\def \ba {\begin{array}}
\def \ea {\end{array}}
\def \bea{\begin{eqnarray}}
\def \eea{\end{eqnarray}}

\def \a {\alpha}
\def \b {\beta}

\def \G {\Gamma}
\def \d {\delta}
\def \D {\Delta}
\def \dg {\dagger}
\def \e {\epsilon}

\def \vp {\varphi}
\def \s {\sigma}

\def \r {\rho}

\def \O {\Omega}

\def \mR {\mathcal R}

\def \mV {\mathcal V}

\def \p {\partial}

\def \mc {\mathcal}

\def \tr {\textrm{tr}}
\def \Tr {{\textrm{Tr}}}

\def \and {{\textrm{and}}}

\begin{document}
	\begin{titlepage}
		
		\title{\textbf {R\'enyi entanglement asymmetry in 1+1-dimensional conformal field theories}}
		 \author{Miao Chen$^{a}$\footnote{miaochen1208@163.com}~, Hui-Huang Chen$^{a}$\footnote{chenhh@jxnu.edu.cn}}
	\maketitle
	\underline{}
	\vspace{-12mm}
	\begin{center}
		{\it
             $^a$College of Physics and Communication Electronics, Jiangxi Normal University,\\ Nanchang 330022, China\\
		}
		\vspace{10mm}
	\end{center}

		\begin{abstract}
        In this paper, we consider the R\'enyi entanglement asymmetry of excited states in the 1+1 dimensional free compact boson conformal field theory (CFT) at equilibrium. We obtain a universal CFT expression written by correlation functions for the charged moments via the replica trick. We provide detailed analytic computations of the second R\'enyi entanglement asymmetry in the free compact boson CFT for excited states $\Psi=V_{\beta}+V_{-\beta}$ and $\Phi=V_{\beta}+J$ with $V_{\beta}$ and $J=i\partial\phi$ being the vertex operator and current operator respectively. We make numerical tests of the universal CFT computations using the XX spin chain model. Taking the non-Hermite fake RDMs into consideration, we propose an effective way to test them numerically, which can be applied to other excited states. The CFT predictions are in perfect agreement with the exact numerical calculations.
		\end{abstract}
		
	\end{titlepage}

	\thispagestyle{empty}

	\newpage

	\tableofcontents
\section{Introduction}
In recent years, there has been a strong research interest in the interplay between the entanglement and symmetries. In condensed matter physics, entanglement is a powerful tool to characterize different phases of matter. In the studies of thermalizations of isolated quantum systems, people found that entanglement is a crucial quantity that characterizes how the subsystem reach equilibrium. \cite{2006Entanglement,2008Area,2009Entanglement,Laflorencie:2015eck}. As the most important concept in modern physics, symmetry and its breaking of a quantum system can lead to a large number of interesting phenomena like ferromagnetism\cite{CK}, superfluidity\cite{2004S} and superconduction. There exist a vast number of references discussing how entanglement decompose under global symmetries both in and out of equilibrium. See for example \cite{2018Symmetry,Cornfeld:2018wbg,Chen:2021pls,Bonsignori:2019naz,Murciano:2019wdl,Chen:2021nma,Horvath:2020vzs,Fraenkel:2019ykl,Murciano:2020vgh,Azses:2020wfx,Parez:2020vsp,Chen:2022gyy,Chen:2023whs,Rath:2022qif,Bertini:2022srv,Fossati:2023zyz} and references therein.  Recently, these theoretical studies have also been confirmed experimentally \cite{2018Probing,Azses:2020tdz,Neven:2021igr,Vitale:2021lds}. 
\par Entanglement entropy or  Von Neumann entropy is the most useful entanglement measure to characterize the bipartite entanglement of a pure state. If we prepare our system in a pure state $\ket{\psi}$, the reduced density matrix (RDM) of the subsystem $A$ is obtained by tracing out degrees of freedom that are not in $A$, i.e. $\rho_A=\tr_{\bar A}\ket{\psi}\bra{\psi}$, where $\bar A$ is the complement of $A$. One can compute the von Neumann entropy from $\Tr\rho_A^{n}$ via the replica trick \cite{2006Entanglement}
\be
S_A\equiv-\Tr(\rho_A\log\rho_A)=\lim_{n\rightarrow 1}S_A^{(n)},
\ee
where $S_A^{(n)}$ is the R\'enyi entropies
\be
S_A^{(n)}=\frac{1}{1-n}\log\Tr\rho_A^{n}.
\ee
\par Recently, the concept of entanglement asymmetry is proposed in \cite{2022E1} as a tool to measure the degree of symmetry breaking in extended quantum systems. Universal formula for matrix product states with finite bond dimension has been obtained in \cite{Capizzi:2023xaf}. In quantum quench problems, the entanglement asymmetry plays a crucial role in exploring whether the initially broken  symmetry can be restored at late times \cite{2022E1,Ares:2023kcz,Bertini:2023ysg,Capizzi:2023qty,Ferro:2023sbn}. Moreover, it helps in finding a kind of quantum Mpemba effect \cite{2022E1,Rylands:2023yzx} and leads to new forms of weak and strong Mpemba effects\cite{Murciano:2023qrv}. Very recently, the microscopic origin of the quantum Mpemba effect in integrable systems has been discussed in \cite{Rylands:2023yzx}. 
\par However, most of the references mentioned above are focused on the time evolution of entanglement asymmetry. As a new and very important quantity, it is valuable to investigate the properties even at equilibrium. In this paper, we are interested in the calculations of entanglement asymmetry of excited states in CFT at equilibrium. To construct states which have non-vanish entanglement asymmetry, we need to consider the superposition states and non-Hermitian terms will occur in the corresponding RDMs. To numerically test our analytical predictions, special attention need to paid to these non-Hermitian terms. Similar quantities also appear in other field of research, see for example \cite{2020Pseudo,2021Aspects,Murciano:2021dga}.
\par The remaining part of this paper is organized as follows. In section 2, we briefly review the concept of entanglement asymmetry and related quantities. In section 3, we discuss how to compute the entanglement asymmetry in conformal field theories (CFT). In section 4, we focus on a particular CFT i.e. the free compact boson theory and make explicit calculations of the R\'enyi entanglement asymmetry for two types of excited states with R\'enyi index $n=2$. In section 5, we numerically test our field theory predictions in the XX spin chain. Finally, we conclude in section 6 and all technical details are presented in four appendices.
\section{Entanglement asymmetry}
\par We prepare an extended quantum system in a pure state $\ket{\psi}$, and divide the system into two spatial regions $A$ and $B$. The reduced density matrix (RDM)  $\rho_A=\Tr_{B}\ket{\psi}\bra{\psi}$ describes the state of the subsystem $A$. We consider a charge operator $Q=Q_A+A_B$ which is the generator of a global $U(1)$ symmetry group. We assume that the state $\ket{\psi}$ is not an eigenstate of $Q$, then$[\rho_A,Q_A]\neq0$. Therefore, $\rho_A$ displays off-diagonal elements in the eigenbasis of the subsystem charge $Q_A$. 
\par
\begin{figure}
	\centering
	{\includegraphics[width=7cm]{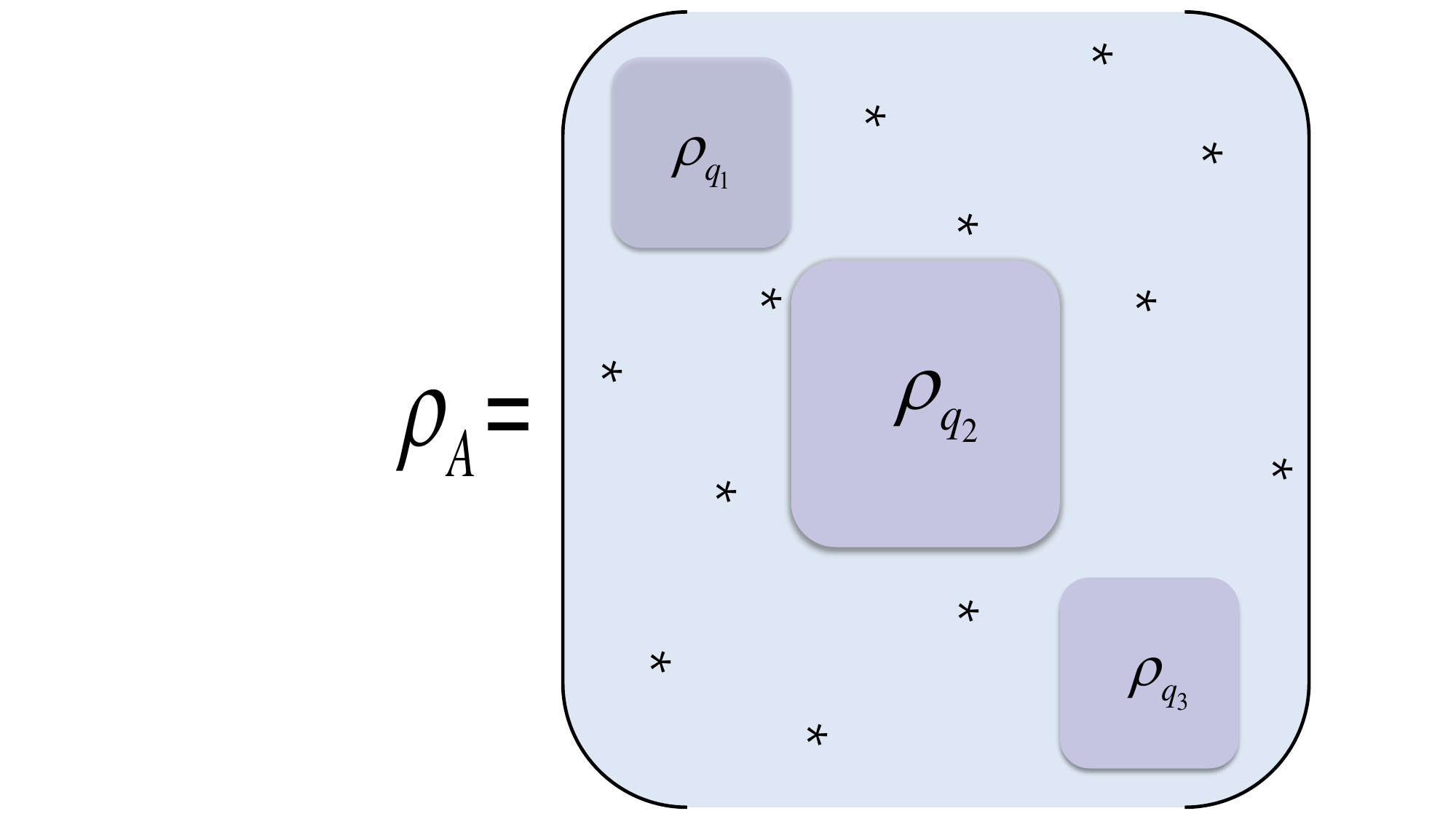}} 
	{\includegraphics[width=7cm]{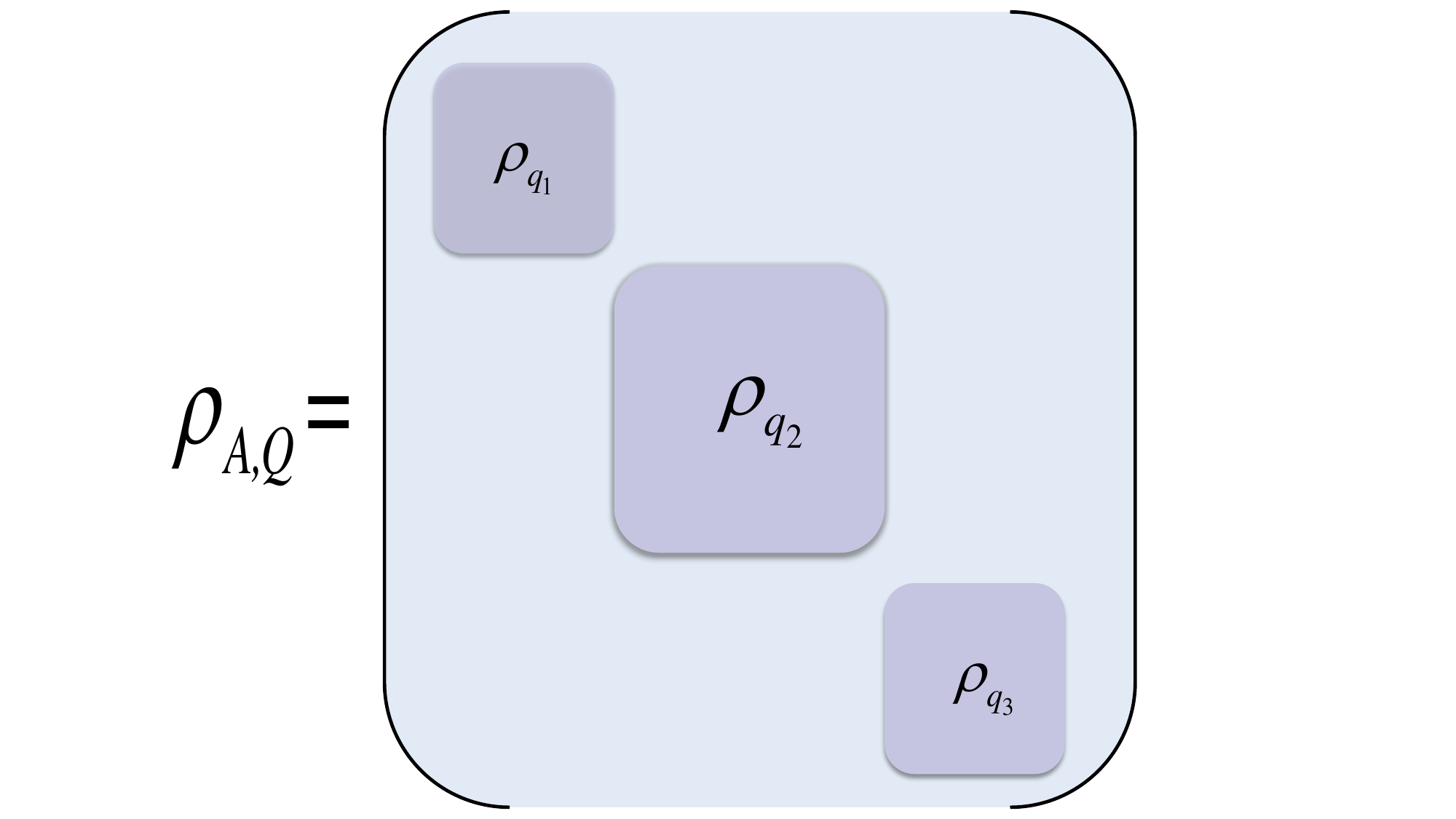}}
	\caption{The schematic comparison of the structure of the density matrices $\rho_A$ and $\rho_{A,Q}$ in the eigenbasis
		of the subsystem charge $Q_A$. The RDM $\rho_A$ contains non-diagonal elements. Instead, $\rho_{A,Q}$ obtained under a projective measurement of $Q_A$, is a block diagonal matrix. The entanglement asymmetry is given by the difference $\Delta S_{A}^{(n)}$ between the entanglement entropies.}
	\label{fig1}
\end{figure}

 For later's convenience, we introduce the quantity $\rho_{A,Q}$ which can be obtained by removing the off-diagonal elements of $\rho_A$,
\be
\rho_{A,Q}=\sum_{q\in\mathbb{Z}}\Pi_q\rho_A\Pi_q,
\ee
where $\Pi_q$ is the projector onto the eigenspace of $Q_A$ with charge $q$. Clearly, $\rho_{A,Q}$ is a block diagonal matrix, i.e. $[\rho_{A,Q},Q_A]=0$. In Fig.~\ref{fig1}, we show the form of $\rho_A$ and $\rho_{A,Q}$ intuitively.
\par To measure the extent to which the symmetry generated by $Q$ is broken in the subsystem $A$, the entanglement asymmetry is introduced in \cite{2022E1} and defined as
\begin{equation}\label{eq:def}
	\Delta S_{A}=S(\rho_{A,Q})-S(\rho_A).
\end{equation}
It can quantity the symmetry breaking at the level of subsystem $A$. Obviously, the entanglement asymmetry is only zero when $\rho_A$ commutes with $Q_A$ i.e. $[\rho_A,Q_A]=0$, and it is always non-negative, $\Delta S_A\geq 0$\cite{Ma:2021zgf}.
\par Using the same strategy to get the Von Neumann entropy, we can define the R\'enyi entanglement asymmetry as
\be\label{san}
	\Delta S_{A}^{(n)}=S^{(n)}(\rho_{A,Q})-S^{(n)}(\rho_A)=\frac{1}{1-n}\left[\log\mathrm{Tr}(\rho_{A, Q}^n)-\log\mathrm{Tr}(\rho_A^n)\right].
\ee
Then $\Delta S_{A}$ can be accessed from $\Delta S_{A}^{(n)}$  by taking the limit $n\rightarrow 1$,
\be
\Delta S_{A}=\lim_{n\rightarrow 1}\Delta S_{A}^{(n)}.
\ee
The R\'enyi entanglement asymmetry are also non-negative, $\Delta S_A^{(n)}\geq 0$, and they vanish only if $[\rho_A,Q_A]=0$\cite{Han:2022phu}.  
Now consider the system with conserved charge $Q=Q_A+Q_B$, with $[\rho_A,Q_A]\neq0$. The density matrix $\rho_{A,Q}$ can be written as block diagonal forms, $\rho_{A,Q}=\sum_{q\in\mathbb{Z}}\Pi_q\rho_A\Pi_q$. Using the integral representation of  $\Pi_q$
\be
\Pi_q=\int_{-\pi}^\pi \frac{{\rm d}\alpha}{2\pi}e^{i\alpha Q_A}e^{-i\alpha q},
\ee
we can obtain the post-measurement density matrix $\rho_{A,Q}$\cite{2022E1} as
\be
\rho_{A, Q}=\int_{-\pi}^\pi \frac{{\rm d}\alpha}{2\pi}e^{-i\alpha Q_A}\rho_A e^{i\alpha Q_A}.
\ee
Then we can write the moments of $\rho_{A,Q}$ as
\be\label{ft}
\Tr(\rho_{A, Q}^n)=\int_{-\pi}^\pi \frac{{\rm d}\alpha_1\dots{\rm d}\alpha_n}{(2\pi)^n} Z_n(\boldsymbol{\alpha}),
\ee
where $\boldsymbol{\alpha}=\{\alpha_1,\dots,\alpha_n\}$ and
\be\label{Za}
Z_n(\boldsymbol{\alpha})=
\mathrm{Tr}\left[\prod_{j=1}^n\rho_A e^{i\alpha_{j,j+1}Q_A}\right],
\ee
with $\alpha_{ij}\equiv\alpha_i-\alpha_j$ and $\alpha_{n+1}=\alpha_1$. $\Tr(\rho_{A, Q}^n)$ are the Fourier transform of the partition function on the $n$-sheet Riemann surface. We can find that if $[\rho_A,Q_A]=0$, $Z_n(\boldsymbol{\alpha})=Z_n(0)$, thus  $\mathrm{Tr}(\rho_{A, Q}^n)=\mathrm{Tr}(\rho_A^n)$ and $\Delta S_{A}^{(n)}=0$. Similar to the case of symmetry resolved entanglement, we call $Z_n(\boldsymbol{\alpha})$ as $charged$ $moments$\cite{2018Symmetry}.
\par The ratio of $\mathrm{Tr}(\rho_{A, Q}^n)$ and $\mathrm{Tr}(\rho_A^n)$ is direct related to the R\'enyi entanglement asymmetry. For further calculation, it's convenient to define  
\be
g_n(A)=\frac{\mathrm{Tr}(\rho_{A, Q}^n)}{\mathrm{Tr}(\rho_A^n)}.
\ee
As a result, according to eq.~(\ref{san}), the R\'enyi entanglement asymmetry $\Delta S_{A}^{(n)}$ can be given by $g_n(A)$,
\be\label{sg}
\Delta S_{A}^{(n)}=\frac{1}{1-n}\log g_n(A).
\ee
\par In terms of eq.~(\ref{ft}), the Fourier transform of the ratio $g_n(A)$ is
\be
g_n(\bm{\a},A)=\frac{\mathrm{Tr}(\prod_{j=1}^n\rho_A e^{i\alpha_{j,j+1}Q_A})}{\mathrm{Tr}(\rho_A^n)},
\ee
which will be used in the latter section.
\par From the analysis in this subsection, to compute the R\'enyi entanglement asymmetry, the most important step is calculating $g_n(\bm{\a},A)$. In next two sections, we will discuss how to compute it in a special CFT, i.e. the free compact boson CFT.
\section{Entanglement asymmetry in CFT}
\subsection{Entanglement of excited states in CFT}
In this section, let's review the replica trick to the entanglement asymmetry in 1+1 dimensional CFT. We consider a periodic 1D system with one spatial dimension, total length $L$, and subsystem A given by the interval $[u,v]$ with length $l=v-u$. It's useful to introduce the dimensionless parameter $x=\frac{v-u}{L}=\frac{l}{L}$, which characterize the size of the subsystem $A$.
\par An infinite cylinder with circumference $L$ can be described as the world sheet of the 1+1 dimensional CFT and parameterized by the introduction of complex coordinate $z$. We are interested in the excited states corresponding to local primary operators
\be 
\ket{\Upsilon} = \Upsilon(-i\infty)\ket{0},
\ee
where $\ket{0}$ is the CFT ground state. In the previous section, we have defined the reduced density matrix (RDM) of subsystem $A$ as $\rho_{\Upsilon}=\Tr_{B}\ket{\Upsilon}\bra{\Upsilon}$. Here we have omit the subscript $A$ and stress on the operator which creates the state. The reduced density matrix of the ground state which corresponds to the identity operator $I$ is written as $\rho_I$. It's convenient to define the ratio
\be\label{G}
G^{\Upsilon}_n(x)\equiv\frac{\Tr(\rho_{\Upsilon}^n)}{\Tr(\rho_I^n)}.
\ee
\par Following the usual strategy, we can obtain $\Tr(\rho_I^n)$ if we sew cyclically $n$ copies of the above cylinders along with the interval $[u,v]$. Different from the ground state case, there are two additional insertions of $\Upsilon(-i\infty) $ and $\Upsilon^{\dagger}(i\infty) $ in the corresponding path-integral representation of the reduced density matrix $\rho_\Upsilon$. In this way, end up with a $n$-sheeted Riemann surface $\mR_n$. Arriving at a $2n$-point  function, $G^{\Upsilon}_n(x)$ is straightforwardly given\cite{Alcaraz:2011tn,Berganza:2011mh,Capizzi:2020jed}
\be\label{G1}
G^{\Upsilon}_n(x) = \frac{\left\langle \displaystyle \prod_{j=1}^{n} \Upsilon(z_j^-) \Upsilon^\dagger(z_j^+) \right\rangle_{\mathcal{R}_n}    }{\langle \Upsilon(z_1^-) \Upsilon^\dagger(z_1^+) \rangle_{\mathcal{R}_1}^n },
\ee
where $z_j^{\mp}$ is points inserting the operators in the $j$-th copy of the system ($j=1,...,n$) in $\mathcal{R}_n$. Obviously, ${\cal R}_1$ is just the cylinder and $G^{\Upsilon}_1(x)=1$, for the normalisation of the involved matrices. 
\par Apply the conformal mapping
\be\label{cfmap}
w(z) = -i\log \left(-\frac{\sin\frac{\pi(z-u)}{L}}{ \sin\frac{\pi(z-v)}{L}}\right)^{1/n},
\ee
to transform the $n$-sheet Riemann surface $\mR_n$ into a single cylinder. The transformation law of a primary field $\mathcal{O}$ is 
\be
\mathcal{O}(w, \bar{w}) = \left( \frac{dz}{dw}\right)^{h_\mathcal{O}}  \left( \frac{d\bar{z}}{d\bar{w}}\right)^{\bar{h}_\mathcal{O}} \mathcal{O}(z,\bar{z})
\ee
with $(h_{\mc{O}},\bar{h}_{\mathcal{O}})$, the conformal weights of $\mathcal{O}$. $G^{\Upsilon}_n(x)$ can be easily accessed from correlation functions on the cylinder under the conformal maps in eq.~(\ref{cfmap})\cite{Berganza:2011mh},
\be\label{Gt}
G^{\Upsilon}_n(x) =  n^{-2n(h_\Upsilon+\bar{h}_\Upsilon)}  \frac{\langle\prod_{j=1}^{n} \Upsilon(w^-_j)\Upsilon^\dagger(w^+_j)\rangle_{\rm cyl}}{\langle\Upsilon(w^-_1)\Upsilon^\dagger(w^+_1)\rangle^n_{\rm cyl}}.
\ee
where $w_j^{\pm}$ are the points corresponding to $z_j^{\pm}$ through the map $w(z)$
\be\label{tkn}
w_j^-=\frac{\pi}{n}((x-1)+2j),\qquad w_j^+=\frac{\pi}{n}(-(x+1)+2j),\qquad j=1,2,\cdots,n.
\ee
\subsection{Entanglement asymmetry in CFT}
In this subsection, let's mimic the idea of the useful strategy before obtaining $g_n(\bm{\a},x)$. In the free compact boson CFT, the $charged$ $moments$ can be studied in a usual way like the process of computing symmetry resolution of entanglement entropy.  
\par Firstly, we consider the ground state case. Let's briefly review the study before, we can regard $\mathrm{Tr}(\rho_A^n e^{i\alpha Q_A})$ as a partition function in the $n$-sheet Riemann surface $\mR_n$ with an inserted Aharonov-Bohm flux $\alpha$. In a similar way, $\mathrm{Tr}(\prod_{j=1}^n\rho_A e^{i\alpha_{j,j+1}Q_A})$ can be seen as a partition function in the $n$-sheet Riemann surface $\mR_n$ with the $j$-th sheet and $j+1$-th sheet inserted Aharonov-Bohm flux $\alpha_{j,j+1}$. It's easy to understand that a twisted boundary condition corresponds to the insertion of a flux. We can introduce a local operator $\mc{V}_{\a_{j,j+1}}$ to encode these twisted boundary conditions\cite{2018Symmetry,Bonsignori:2019naz}. If $A$ is an interval $[u,v]$, we can obtain the following relation\cite{Capizzi:2020jed}
\be\label{va}
e^{i\a_{j,j+1} Q_A}=\mc{V}_{\a_{j,j+1}}(u,0)\mc{V}_{-\a_{j,j+1}}(v,0).
\ee
Moreover, one can precisely identity
\be\label{ga2}
\mathrm{Tr}(\prod_{j=1}^n\rho_I e^{i\alpha_{j,j+1}Q_A})=\langle \prod_{j=1}^n e^{i\alpha_{j,j+1}Q_A}\rangle_{\mc{R}_n}=\langle\prod_{j=1}^n \mc{V}_{\a_{j,j+1}}(u,0) \mc{V}_{-\a_{j,j+1}}(v,0)\rangle_{\mc{R}_n}.
\ee 
\newline Refer to the result in the last section  eq.~(\ref{G1}), we similarly have
\be\label{g0}
\mathrm{Tr}(\prod_{j=1}^n\rho_{\Upsilon} e^{i\alpha_{j,j+1}Q_A})=\langle\prod_{j=1}^n \mc{V}_{\a_{j,j+1}}(u,0) \mc{V}_{-\a_{j,j+1}}(v,0)\Upsilon(z_j^-) \Upsilon^\dagger(z_j^+)\rangle_{\mc{R}_n}.
\ee
Therefore, $g^{\Upsilon}_n(\bm{\a},x)$ can be computed as
\be\label{ga}
g^{\Upsilon}_n(\bm{\a},x)
=\frac{\langle\prod_{j=1}^n \mc{V}_{\a_{j,j+1}}(u,0) \mc{V}_{-\a_{j,j+1}}(v,0)\psi(z_j^-) \psi^\dagger(z_j^+)\rangle_{\mc{R}_n}}{\langle\prod_{j=1}^n \Upsilon(z_j^-) \Upsilon^\dagger(z_j^+)\rangle_{\mc{R}_n}}.
\ee
\par Through the conformal transformation defined in eq.~(\ref{cfmap}), all the correlation functions in eq.~(\ref{ga}) can be mapped to correlators on the cylinder. Furthermore, all powers of $(\frac{dz}{dw})$ cancel out in this mapping. Consequently, we can write $g_n(\bm{\a},x)$ as 
\be
g^{\Upsilon}_n(\bm{\a},x)=\frac{\langle\prod_{j=1}^n \mc{V}_{\a_{j,j+1}}(i\infty) \mc{V}_{-\a_{j,j+1}}(-i\infty)\Upsilon(w_j^-) \Upsilon^\dagger(w_j^+)\rangle_{\rm_{cyl}}}{\langle\prod_{j=1}^n \Upsilon(w_j^-) \Upsilon^\dagger(w_j^+)\rangle_{\rm_{cyl}}}.
\ee
It's a universal CFT expression written by correlation functions for the charged moments using the replica trick, which is the bridge to the entanglement asymmetry in CFT. We will use the equation above for two types of excited states in CFT. Detailed analytic computations will be discussed in the next section.

\section{Excited states in the free compact boson CFT}
In this section and the following part, we will focus on the entanglement asymmetry in the free compact boson CFT. 
\par The theory of free compact bosonic field $\vp(z,\bar{z})$ with Euclidean action 
\be
\mc{A}[\vp]=\frac{1}{8\pi}\int dzd\bar{z}\p_z\varphi\p_{\bar{z}}\vp
\ee
 is a CFT with a central charge $c=1$. This theory has two types of primary fields and the first type is the vertex operators
\be
V_{\a,\bar{\a}}=:e^{i(\a\phi+\bar{\a}\bar{\phi})}:
\ee
where $\phi,\bar{\phi}$ are chiral and anti-chiral portions of the bosonic field: $\vp(z,\bar{z})=\phi(z)+\bar{\phi}(\bar{z})$, with the conformal weight $(h, \bar{h}) = \left(\frac{\a^2}{2},\frac{\bar{\a}^2}{2} \right)$ consisting of the holomorphic 
and the anti-holomorphic sectors. Another primary operator is the current operator or the derivative operator $J=i\p\phi$. For computing easily, we assume holomorphic field $\bar\phi=0$.
\par Considering that the conserved current is proportional to $\partial_x \phi$, the charge operator in the interval $A$ is
\be 
Q_A = \frac{1}{2\pi}\int_A dx \ \partial_x \varphi = \frac{1}{2\pi}(\varphi(v)-\varphi(u)).
\ee 
Under the inspection of eq.~(\ref{va}), the local operator mentioned above $\mc{V}_{\a_{j,j+1}}$ is implemented by the vertex operator 
\be 
\mathcal{V}_{\a_{j,j+1}} =V_{\frac{\a_{j,j+1}}{2\pi}}\equiv e^{\frac{i\a_{j,j+1}}{2\pi} \varphi},
\ee
with the conformal weight $(h_{\a_{j,j+1}}, \bar{h}_{{\a_{j,j+1}}}) = \left(\frac{1}{2}(\frac{\a_{j,j+1}}{2\pi})^2,\frac{1}{2}(\frac{\a_{j,j+1}}{2\pi})^2 \right)$.
\par  During the calculation, it was found that we can't achieve $\mathrm{Tr}(\prod_{j=1}^n\rho_A e^{i\alpha_{j,j+1}Q_A})$ straightly. For the total entanglement, the ratio of  moments eq.~(\ref{G}) is universal and can be calculated in CFT without any input from the model\cite{Alcaraz:2011tn}.  Inspired by it, we can define the following ratio of charged moments
\be  
G^{\Upsilon}_n(\boldsymbol{\alpha},x)=\frac{\mathrm{Tr}(\prod_{j=1}^n\rho_{\Upsilon}e^{i\alpha_{j,j+1}Q_A})}{\mathrm{Tr}(\prod_{j=1}^n\rho_I e^{i\alpha_{j,j+1}Q_A})},
\ee
which is also independent and universal of any microscopic details. Applying the same technique as above, we can rewrite it as
 \be
G^{\Upsilon}_n(\boldsymbol{\alpha},x)=\frac{\langle\prod_{j=1}^n \mc{V}_{\a_{j,j+1}}(i\infty) \mc{V}_{-\a_{j,j+1}}(-i\infty)\Upsilon(w_j^-) \Upsilon^\dagger(w_j^+)\rangle_{\rm_{cyl}}}{\langle\prod_{j=1}^n \mc{V}_{\a_{j,j+1}}(i\infty) \mc{V}_{-\a_{j,j+1}}(-i\infty)\rangle_{\rm_{cyl}}},
 \ee
 which will be computed in the following part. Notice that at $\bm{\a}=0$, $G^{\Upsilon}_n(0,x)=G^{\Upsilon}_n(x)$. The observation suggests to define another ratio
\be
\begin{split}
&\frac{G^{\Upsilon}_n(\boldsymbol{\alpha},x)}{G^{\Upsilon}_n(0,x)}=\frac{\mathrm{Tr}(\prod_{j=1}^n\rho_{\Upsilon}e^{i\alpha_{j,j+1}Q_A})\mathrm{Tr}(\rho_I^n)}{\mathrm{Tr}(\prod_{j=1}^n\rho_I e^{i\alpha_{j,j+1}Q_A})\mathrm{Tr}(\rho_{\Upsilon}^n)}=\frac{g_n^{\Upsilon}(\bm{\a},x)}{g_n^{I}(\bm{\a},x)}\\&=\frac{\langle\prod_{j=1}^n \mc{V}_{\a_{j,j+1}}(i\infty) \mc{V}_{-\a_{j,j+1}}(-i\infty)\Upsilon(w_j^-) \Upsilon^\dagger(w_j^+)\rangle_{\rm_{cyl}}}{\langle\prod_{j=1}^n \mc{V}_{\a_{j,j+1}}(i\infty) \mc{V}_{-\a_{j,j+1}}(-i\infty)\rangle_{\rm_{cyl}}\langle\prod_{j=1}^n \Upsilon(w_j^-) \Upsilon^\dagger(w_j^+) \rangle_{\rm_{cyl}}}
\end{split}
\ee
\par From the analysis above, it's of most importance to get $G^{\Upsilon}_n(\boldsymbol{\alpha},x)$. In this section, we will discuss how to get $G^{\Upsilon}_n(\boldsymbol{\alpha},x)$ with two types of excited states in the free compact boson CFT.
\par Special attention needs to be paid to the property of the excited state. If the excited state $\ket{\Upsilon}$ is an eigenstate of $Q$, then $[\rho_A,Q_A]=0$. Thus the a state like $\ket{V_{\b}}$ and $\ket{J}$ will lead to a vanishing entanglement asymmetry. Excited states we will consider must not being eigenstates of $Q$. These state satisfy $[\rho_A,Q_A]\neq0$, so their entanglement asymmetries aren't zero.
\par Regarding with the above discussions, we will calculate $g^{\Upsilon}_2(x)$ for two kinds of excited states, i.e. $\ket{\Psi}=\ket{V_\b}+\ket{V_{-\b}}$ and $\ket{\Phi}=\ket{V_{\b}}+\ket{J}$. It is not hard to extend our results to other kinds of excited states.

\subsection{Excited state $\textbf{I}$: $\ket{\Psi}=\ket{V_\b}+\ket{V_{-\b}}$}
In this subsection, we will focus on the computation of $g^{\Psi}_2(x)$, with $\ket{\Psi}=\ket{V_\b}+\ket{V_{-\b}}$. which is the key ingredient of the R\'enyi entanglement asymmetry.
\be\label{Psi}
\begin{split}
	\ket{\Psi}\bra{\Psi}&=(\ket{V_\b}+\ket{V_{-\b}})(\bra{V_\b}+\bra{V_{-\b}})\\
	&=\ket{V_\b}\bra{V_\b}+\ket{V_\b}\bra{V_{-\b}}+\ket{V_{-\b}}\bra{V_\b}+\ket{V_{-\b}}\bra{V_{-\b}}.
\end{split}
\ee
\par We know that the correlation function of vertex operators $\langle V_{\a_1}V_{\a_2}...V_{\a_N}\rangle$ isn't zero, only if $\sum_{i=1}^{N}\a_i=0$. Taking this neutral condition into account, we find there are only 6 terms contribute
\be\label{G2psi}
\begin{split}
	G^{\Psi}_2(\boldsymbol{\alpha},x)&=F_{\b,\b,-\b,-\b}+F_{-\b,-\b,\b,\b}+F_{\b,-\b,\b,-\b}\\
	&+F_{-\b,\b,-\b,\b}+F_{\b,-\b,-\b,\b}+F_{-\b,\b,\b,-\b}
\end{split}
\ee\label{f1}
where 
\be
F_{\b,\b,-\b,-\b}=\frac{\langle\mV_{\a}(i\infty) \mV_{-\a}(-i\infty)V_{\b}(w_1^-) V_{\b}(w_1^+)\mV_{-\a}(i\infty) \mV_{\a}(-i\infty)V_{-\b}(w_2^-) V_{-\b}(w_2^+)\rangle_{\rm_{cyl}}}{\langle \mV_{\a}(i\infty)\mV_{-\a}(-i\infty)\mV_{-\a}(i\infty)\mV_{\a}(-i\infty)\rangle_{\rm_{cyl}}}.
\ee
The subscript of the function $F$ indicates the order of the corresponding vertex operator appearing in the correlator. The explicit expressions of other terms in eq.~(\ref{G2psi}) can be written down in a similar way. Here and in the following, we are using the notation $\a\equiv \a_{1,2}=-\a_{2,1}$.
\par We know that the correlation functions of an arbitrary number of vertex operators on the cylinder can be given by elementary methods\cite{DiFrancesco:1997nk}
\be\label{vv}
\langle V_{\alpha_1}(z_1)...V_{\alpha_n}(z_n)\rangle_{\rm cyl} = \prod_{i<j} \big(2\sin \frac{z_i-z_j}{2}\big)^{\alpha_i\alpha_j}.
\ee
In the following, we plan to calculate $G^{\Psi}_2(\boldsymbol{\alpha},x)$, referring to this equation. We find that the computation is complicated but straightforward and the details of the calculation are presented in appendix \ref{appenA}. Finally, the R\'enyi entanglement asymmetry with index $n=2$ is
\be
\D S_A^{(2)}(\rho_{\Psi})=-\log\frac{a_1\sin^2{(2\pi\b)}+2b_1\pi^2\b^2}{2\pi^2\b^2(2a_1+b_1)}.
\ee
with $a_1=(2\cot{(\frac{\pi x}{2})})^{-2\beta^2}$ and $b_1=2(\sin{\pi x})^{-2\beta^2}+2(2\tan{\frac{\pi x}{2}})^{-2\beta^2}$. 
\subsection{Excited state $\textbf{II}$: $\ket{\Phi}=\ket{V_{\b}}+\ket{J}$}
In this subsection, we consider another type of excited state, which is the superposition of the vertex operator and the derivative operator i.e. $\ket{\Phi}=\ket{V_{\b}}+\ket{J}$. In contrast to the last subsection, a similar but different way will be used. Now we have
 \be\label{Phi}
 \begin{split}
 	\ket{\Phi}\bra{\Phi}&=(\ket{V_\b}+\ket{J})(\bra{V_\b}+\bra{J})\\
 	&=\ket{V_\b}\bra{V_\b}+\ket{V_\b}\bra{J}+\ket{J}\bra{V_\b}+\ket{J}\bra{J}.
 \end{split}
 \ee
 \par For simplicity, we will still focus on the case $n=2$ and higher $n$ can be computed similarly. We have
 \be\label{G2phi}
 \begin{split}
 G^{\Phi}_2(\boldsymbol{\alpha},x)&=F_{\b,-\b,\b,-\b}
+F_{J,-\b,\b,J^{\dg}}+F_{\b,J^{\dg},J,-\b}\\&+F_{J,J^{\dg},\b,-\b}+F_{\b,-\b,J,J^{\dg}}+F_{J,J^{\dg},J,J^{\dg}}.
 \end{split}
 \ee
 where
 \be
 F_{\b,-\b,J,J^{\dg}}=\frac{\langle\mV_{\a}(i\infty)\mV_{-\a}(-i\infty)V_{\b}(w_1^-)V_{-\b}(w_1^+)\mV_{-\a}(i\infty)\mV_{\a}(-i\infty)J(w_2^-) J^{\dg}(w_2^+)\rangle_{\rm_{cyl}}}{\langle \mV_{\a}(i\infty)\mV_{-\a}(-i\infty)\mV_{-\a}(i\infty)\mV_{\a}(-i\infty)\rangle_{\rm_{cyl}}}.
 \ee
 According to the rules of our notation, it's easy to write down the explicit expression of other terms in the equation above.
\par Since the direct calculation of the correlation functions of vertex and derivative operators is usually not an easy task, the standard and useful trick is\cite{Capizzi:2020jed} 
\be\label{dtr}
J(z)=i\partial\phi(z)=\frac{1}{\e}\partial_z V_\e(z)|_{\e=0}
\ee
We can use this to calculate the various correlation functions. The calculation is easy and straightforward and we report the final results in appendix \ref{appenB}.
Adding these results together, one can get $G^{\Phi}_2(\boldsymbol{\alpha},x)$ and hence $g_2^{\Phi}(x)$. Finally, the R\'enyi entanglement asymmetry $\D S^{(2)}_A(\rho_{\Phi})$ is given by
\be
\D S^{(2)}_A(\rho_{\Phi})=-\log\frac{2a_2\sin^2{(\pi\b)}+b_2\pi^2\b^2}{\pi^2\b^2(2a_2+b_2)}
\ee
with 
\be\label{a2}
a_2=2^{-2-\b^2}(-\sec(\frac{\pi x}{2}))^{\b^2}(\b^2\cot^2(\frac{\pi x}{2})+\sec^2(\frac{\pi x}{2}))
\ee
and 
\be\label{b2}
\begin{split}
b_2=&2^{-2-\b^2}(\csc(\frac{\pi x}{2}))^{\b^2}(\csc^2(\frac{\pi x}{2})+\b^2\tan^2(\frac{\pi x}{2}))+2^{-3-\b^2}(2-3\b^2+\b^2(\cos(\pi x)+\\&2\sec^2(\frac{\pi x}{2}))(\sin(\frac{\pi x}{2}))^{-2-\b^2}+(\csc(\pi x))^{2\b^2}+2^{-6}(7+\cos(2\pi x))^2\csc^4(\pi x)
\end{split}
\ee
\par In this section, we give the strategy for how to compute the R\'enyi entanglement asymmetry $\D S^{(2)}_A$ and worked out the exact results for $n=2$. Moreover, it's not too hard to extend our result for $n>2$ and for other types of excited states with non-vanishing R\'enyi entanglement asymmetry.
\section{Numerical tests}
In this section, we will make some numerical tests of the universal CFT computations obtained in previous sections\cite{Chung:2001zz,2002Calculation,2009Reduced}. Based on the calculations above, we plan to numerically calculate the individual terms occur in the original expressions of $G_2^{\Psi}$ and $G_2^{\Phi}$ (c.f. eq.~(\ref{G2psi}) and eq.~(\ref{G2phi})). We will take the XX spin chain model with periodic boundary conditions as an concrete lattice realization of our free compact boson CFT.  As is well known, the XX spin chain is described by the following Hamiltonian\cite{2001Book}
\be
H=-\frac{1}{4}\sum_{j=1}^L(\s^x_j\s^x_{j+1}+\s^y_j\s^y_{j+1}-h\s^z_j),
\ee
where $\s^{x,y,z}_j$ are the Pauli matrices acting on the $j$-th site. After a Jordan-Wigner transformation and Fourier transformation, it can be diagonalized and the eigenstates of the Hamilton are described by a set of momenta $K$, $\ket{K}=\prod_{k\in K}b_k^{\dg}\ket{0}$. See appendix \ref{appenC} for details.
\begin{figure}
	\centering
	\subfloat
	{\includegraphics[width=13cm,height=5cm]{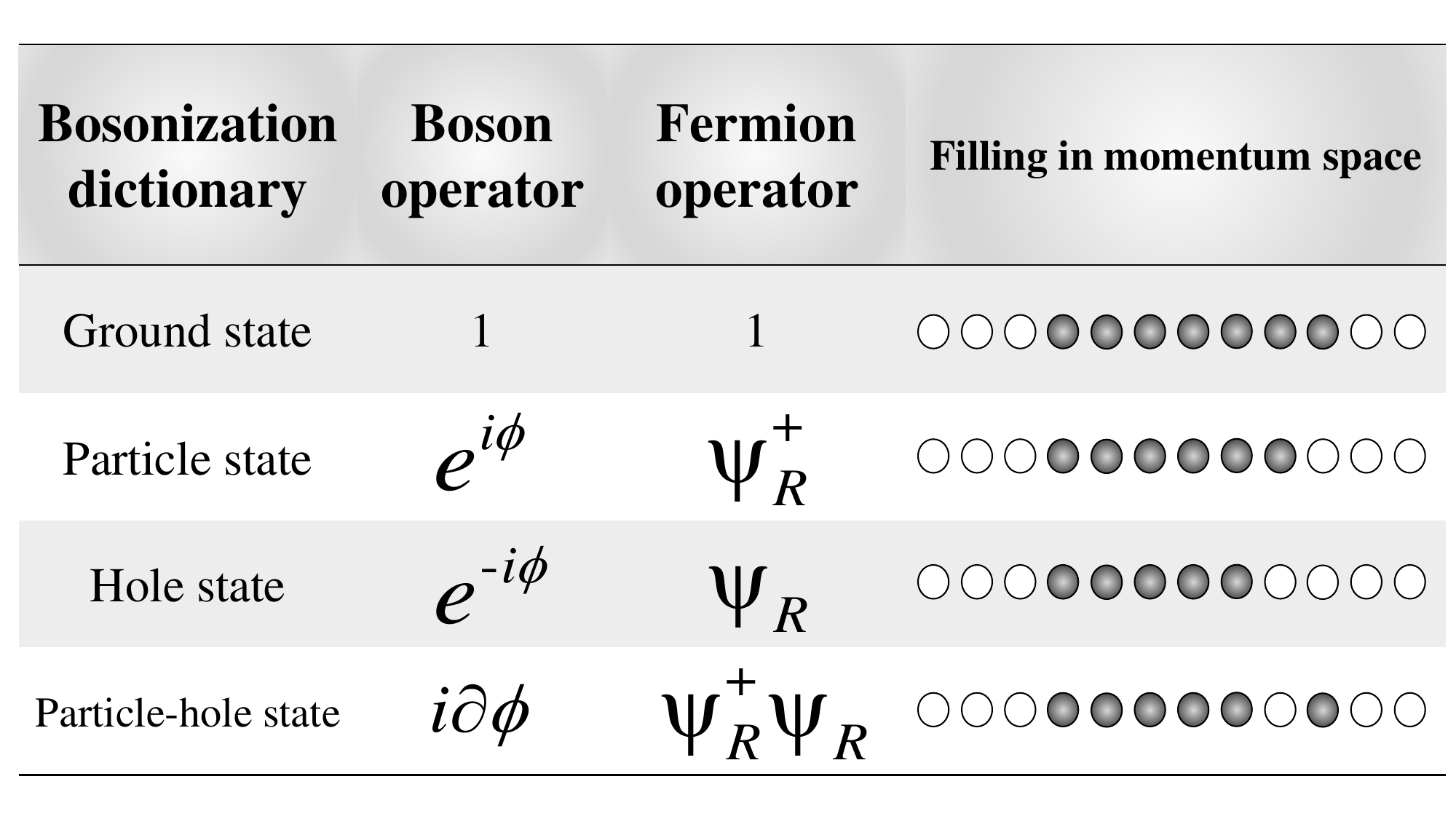}} 
	\caption{Bosonization dictionary for some low-energy excitations of a free-fermion chain with the Hamiltonian $H=-\frac{1}{4}\sum_{j=1}^L(\s^x_j\s^x_{j+1}+\s^y_j\s^y_{j+1}-h\s^z_j)$.}
	\label{fig2}
\end{figure}
\par If we prepare the spin chain in the state $\rho=\ket{K}\bra{K}$ and the consider the case where the subsystem $A$ consists of $l$ continuous sites, we write the Majorana correlation matrix as
\be\label{aa}
\Tr(\rho a_ia_j)=\d_{ij}+\G^K_{ij},
\ee
with $\G^{K}\in\mc{M}_{2l}(\mathbb{C})$. $\G^K$ is a block matrix with elements given by 
\be\label{Gamma}
\G^K_{ij}=T^{K}_{j-i},
\qquad T^K_{n}=\begin{pmatrix}
	f^K_{n}&g^K_{n}\\-g^K_{n}&f^K_{n}
\end{pmatrix},
\ee
In eq.~(\ref{G2psi}) and eq.~(\ref{G2phi}), we find that there are Hermite and non-Hermite terms. For Hermite terms, $f^K_{n}$ and $g^K_{n}$ have the following form
\be
\begin{split}
	&f^K_{n}=\frac{1}{L}\sum_{k\in K}e^{i\phi_kn}-\frac1L\sum_{k\in K}e^{-i\phi_kn},\\
	&g_{n}^K=-\frac{i}{L}\sum_{k\in K}e^{-i\phi_kn}+\frac{i}{L}\sum_{k\notin K}e^{i\phi_kn}.
\end{split}
\label{fg}
\ee
Moreover, their CFT predictions have been tested with the concrete lattice calculations in the XX spin chain and performed well in \cite{Chen:2021pls}. Therefore, all we need is to check whether the CFT results of the non-Hermite parts are in perfect agreement with the exact numerical calculations. 
\par The low-lying states are excitations of holes and particles below or above the Fermi Sea. The correspondences of the vertex and derivative operators will be found by recalling the bosonization dictionary, which helps us a lot in
providing the numerical tests in the two excited states discussed in the previous section. Some examples of the correspondence between CFT operators and the low-energy excitations in XX spin chain are shown in Fig.~\ref{fig2}.
\subsection{Excited state $\textbf{I}$: $\ket{\Psi}=\ket{V_{\b}}+\ket{V_{-\b}}$}
For simplicity, we assume that $\b=1$. The vertex operator $V_{-1}=e^{-i\phi}$  corresponds to a hole excitation and $V_{1}=e^{i\phi}$ corresponds to a particle excitation, at the Fermi momentum\cite{Alcaraz:2011tn}. Because of $n_F=\frac{L}{2}$ and its corresponding $\{\pm\frac{1}{2},\pm\frac{3}{2},\cdots\pm\frac{n_F-1}{2}\}$, the relations of these two states is 
\be\label{1-1}
|V_1\rangle=b_q^{\dg}b_{q'}^{\dg}|V_{-1}\rangle
\quad\quad
q=\frac{L}{4}-\frac{1}{2}, q'=\frac{L}{4}+\frac{1}{2}.
\ee
\par Non-Hermite terms $\ket{V_\b}\bra{V_{-\b}}$ and $\ket{V_{-\b}}\bra{V_{\b}}$ apearing in eq.~(\ref{G2psi}) cannot be viewed as some density matrices, since their traces are zero.  To deal with this kind of object, we introduce the operator $S$ satisfying $S^2=I$ and $\langle V_{1}|S|V_{-1} \rangle \neq 0$ to define a fake density matrix
\be
\rho^S=\frac{S\ket{V_{-1}}\bra{V_1}}{\bra{V_1}S\ket{V_{-1}}}
\ee
The corresponding Majorana matrix can be defined in the following way\cite{2010Entanglement}
\be\label{aa}
\G^S_{ij}=\Tr(\rho^S a_ia_j)-\d_{ij}=\frac{\langle V_{1}|Sa_ia_j|V_{-1}\rangle}{\langle V_{1}|S|V_{-1}\rangle}-\d_{ij},
\ee
\par For different non-Hermitian terms, we should construct an appropriate operator $S$, with the conditions mentioned above being satisfied. For the term $\ket{V_{-1}}\bra{V_{1}}$, we could choose the $S$ operator as
\be\label{s}
 S=\s_n^x\s_{n+1}^y=-ia_{2n-1}a_{2n+1}=i(c_n-c_n^+)(c_{n+1}-c_{n+1}^{\dg}).
\ee
Other forms of $S$ fulfilling the conditions mentioned before are also workable. Refer to eq.~(\ref{s}), we have
\be
\begin{split}
	\langle V_{1}|S|V_{-1}\rangle=i\langle c_n^{\dg}c_{n+1}^{\dg}b_qb_{q'}\rangle_{V_1}&=\frac{i}{L}\sum_{k,k'\in\O}e^{-i\phi_kn-i\phi_{k'}(n+1)}\langle b_k^{\dg}b_{k'}^{\dg}b_qb_{q'}\rangle_{V_1}\\&=\frac{i}{L}[e^{-i\phi_qn-i\phi_{q'}(n+1)}-e^{-i\phi_{q'}n-i\phi_q(n+1)}]\equiv if_0(n,n+1).
\end{split}
\ee
Where $c_n=\frac{1}{\sqrt{L}}\sum_{k\in\O}b_ke^{i\phi_k n}$ with $\phi_k=\frac{2\pi k}{L}$. 
\par Similarly, the Majorana matrix $\G^S$ is a block matrix with elements given by
\be
\G^S_{r,s}=T^S_{r,s},\qquad 
 T^S_{r,s}=\begin{pmatrix}
	f_1(r,s)&g_1(r,s)\\
	g_2(r,s)&f_2(r,s)\end{pmatrix}. 
\ee
The explicit form of the functions $f_1,f_2,g_1$ and $g_2$ can be found in appendix \ref{appenD}. 
\par
\begin{figure}
	\centering
	\subfloat
	{\includegraphics[width=7cm]{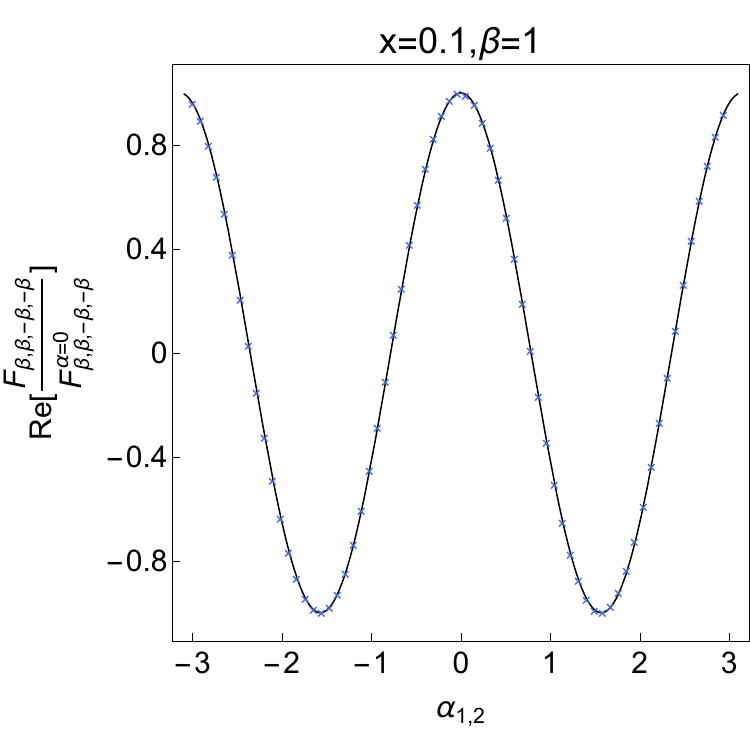}} \quad\quad
	{\includegraphics[width=7cm]{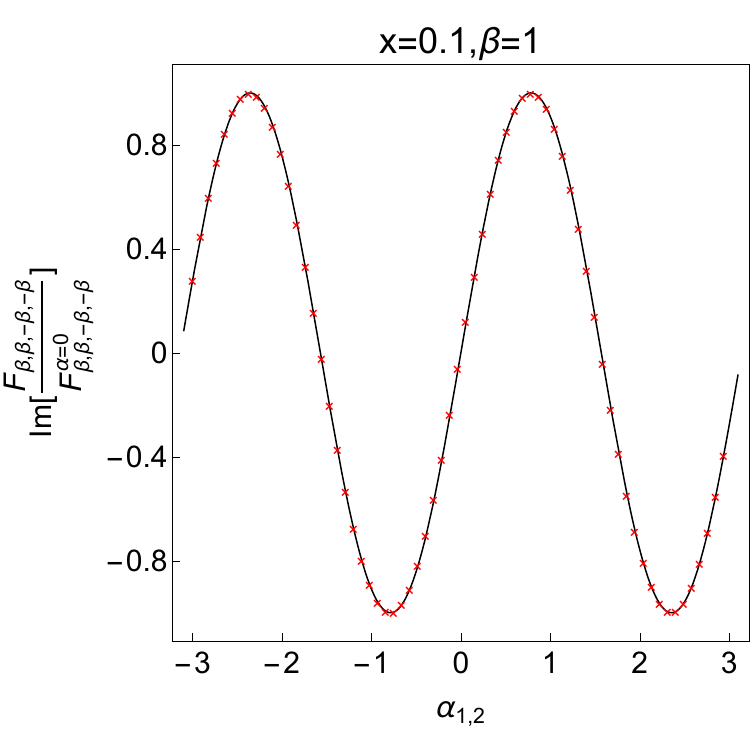}}
	\caption{Numerical result of ${F_{\b,\b,-\b,-\b}}/{F^{\a=0}_{\b,\b,-\b,-\b}}$ in the XX spin chain. The full lines are the CFT results in eq.~(\ref{fb1}). Here we consider $x=0.1, \b=1, n=2$ with total length $L=1000$ and subsystem size $l=100$. As shown in the figure, the agreement is fairly well for $\a_{1,2}\in[-\pi,\pi) $}.
	\label{fig3}
\end{figure}

 We numerically compute the $\a$-dependence of the normalized $F_{\b,\b,-\b,-\b}$ for $\b=1$, i.e.\\
 ${F_{1,1,-1,-1}}/{F^{\a=0}_{1,1,-1,-1}}$. The numerical data are shown in Fig.~\ref{fig3}. As a comparison, the following analytical predictions
\be\label{fb1}
\frac{F_{1,1,-1,-1}}{F^{\a=0}_{1,1,-1,-1}}=e^{2i\alpha}=\cos{(2\alpha)}+i\sin{(2\alpha)},
\ee 
are also plot as the full line in Fig.~\ref{fig3}.

\par
\par Our numerical data should converge to the CFT result in eq.~(\ref{fb1}), in the limit $L\to\infty$. As shown in the figure, the agreement between CFT prediction and numerical data is extremely excellent for all $\a\in[-\pi,\pi)$. Although not showed here, we find that for $\b=-1$, the CFT prediction and the numerical result also match very well.
\subsection{Excited state $\textbf{II}$: $\ket{\Phi}=\ket{V_{\b}}+\ket{J}$}
In this subsection, we will discuss the numerical test in the excited state $\ket{\psi}=\ket{V_{\b}}+\ket{J}$. As discussed in the previous subsection, for non-Hermitian terms like $\ket{V_{\b}}\bra{J}$, we should choose a appropriate operator $S$ to define the corresponding fake density matrix.
For simplicity, we take $\b=1$ again. For the non-Hermite $\ket{V_1}\bra{J}$ and $\ket{J}\bra{V_1}$, we should find out the operator $S$ . The first step is to find the relationship between $|V_1\rangle$ and $|J\rangle$. The derivative operator $J$ corresponds to a right-moving particle-hole excitation so that we can write\cite{Alcaraz:2011tn}
\be
|V_1\rangle=b_q^{\dg}|J\rangle,
\qquad 
q=\frac{L}{4}-\frac{1}{2}
\ee
In this situation, we could choose $S$ as
\be
S=a_{2n-1}=i(c_n-c_n^{\dg})
\ee
The corresponding Majorana matrix is
\be
\G_{ij}^S=\frac{\langle V_1|Sa_ia_j|J\rangle}{\langle V_{1}|S|J\rangle}-\d_{ij}
\ee
where
\be
\langle V_1|S|J \rangle=-i\sum_{k\in\O} e^{-i\phi_kn}\langle b_k^{\dg}b_q\rangle_{V_1}=-\frac{i}{\sqrt{L}}e^{-i\phi_q n}\equiv-ig_0(n)
\ee
\par In this case, the Majorana matrix $\G^S$ is a block matrix with elements given by
\be
\G^S_{r,s}=\begin{pmatrix}
	\tilde{f}_1(r,s)&\tilde{g}_1(r,s)\\
	\tilde{g}_2(r,s)&\tilde{f}_2(r,s)\end{pmatrix}. 
\ee 
The explicit form of $\G_{ij}^S$ can be found in appendix \ref{appenD}. 
\begin{figure}
	\centering
	\subfloat
	{\includegraphics[width=7cm]{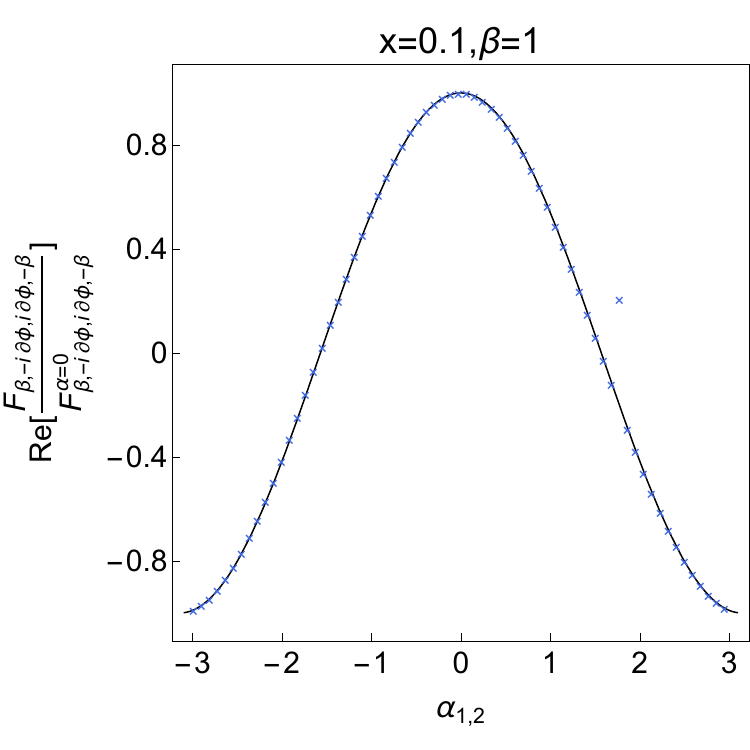}} \quad\quad
	{\includegraphics[width=7cm]{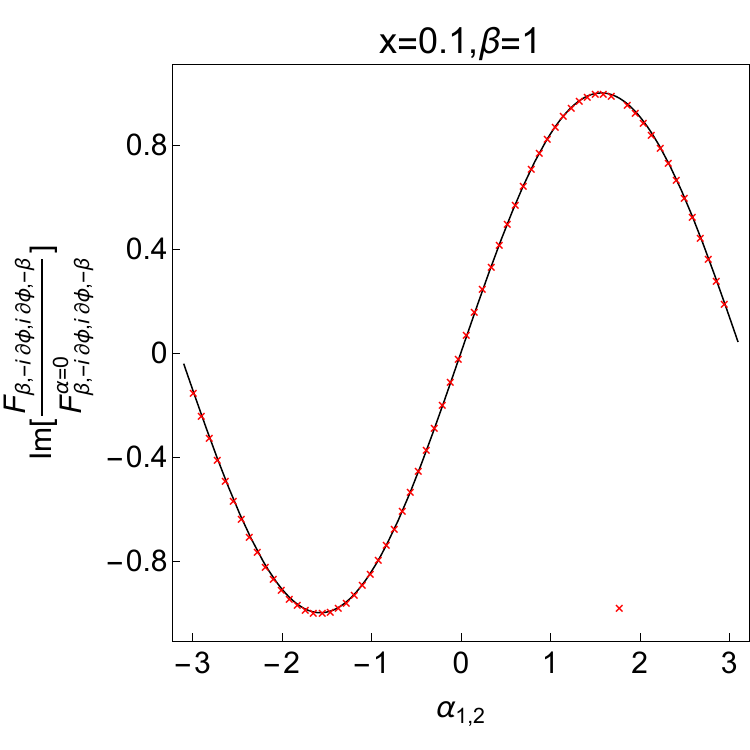}}
	\caption{Numerical result of ${F_{\b,-i\partial\phi,i\partial\phi,-\b}}/{F^{\a=0}_{\b,-i\partial\phi,i\partial\phi,-\b}}$ in the XX spin chain. The full lines are the CFT results in eq.~(\ref{fbfb1}). Here we consider $x=0.1, \b=1, n=2$ with total length $L=1000$ and subsystem size $l=100$. Again, the CFT prediction and the numerical result also match extremely well for $\a_{1,2}\in[-\pi,\pi)$.}
	\label{fig4}
\end{figure}
\par The CFT predictions of the normalized $F_{\b,J^{\dg},J,-\b}$ for $\b=1$, i.e. ${F_{1,J^{\dg},J,-1}}/{F^{\a=0}_{1,J^{\dg},J,-1}}$ is
\be\label{fbfb1}
\frac{F_{1,J^{\dg},J,-1}}{F^{\a=0}_{1,J^{\dg},J,-1}}=e^{i\a}=\cos{\alpha}+i\sin{\alpha},
\ee
We numerically compute ${F_{1,J^{\dg},J,-1}}/{F^{\a=0}_{1,J^{\dg},J,-1}}$  and the data are shown in Fig.~\ref{fig4} as dots. As shown in this figure, the agreement between CFT prediction and numerical result is perfect for all $\a\in[-\pi,\pi)$. 
\section{Conclusion}
In this manuscript, we study the R\'enyi entanglement asymmetry of two excited states in the free compact boson CFT and the underlying lattice model. 
\par Using the replica method, we obtain a universal CFT expression written by the correlation functions for the charge moments. We mention that the entanglement asymmetry of an eigenstate of $Q$ will vanish. Thus we construct two types of excited states: $\ket{\Psi}=\ket{V_\b}+\ket{V_{-\b}}$ and $\ket{\Phi}=\ket{V_{\b}}+\ket{J}$, which are not the eigenstates of the charge.  Considering the characteristics of vertex and current operators, we are able to calculate the correlation functions involving vertex and current operators. As an concrete example, we obtained the exact results for the R\'enyi index $n=2$ from the CFT computation. 
\par The numerical method of computing the charged moments with non-Hermitian fake RDMs has never been studied before. In this paper, we propose a efficient way to treat them numerically. The CFT predictions are in perfect agreement with the exact numerical calculations, which confirms our numerical method.
\par In this paper, we only consider R\'enyi entanglement asymmetry with index $n=2$, it would be very interesting to derive a formula for general $n$ and analytical continue to $n=1$ to obtain the entanglement asymmetry. It will also be interesting the consider other kinds of excited states and the content of operator $S$ should be change to adapt this modification. 
\section*{Acknowledgments}
\begin{appendix}
	\section{Correlation functions of vertex  operators }\label{appenA}
The  most important basis of the appendix is eq.~(\ref{vv})
\be\label{vv}
\langle V_{\alpha_1}(z_1)...V_{\alpha_n}(z_n)\rangle_{\rm cyl} = \prod_{i<j} \big(2\sin \frac{z_i-z_j}{2}\big)^{\alpha_i\alpha_j}.
\ee
\par Using the above formula, it's straightforward to obtain
\be
\begin{split}
&F_{\b,\b,-\b,-\b}=(2\cot{(\frac{\pi x}{2})})^{-2\beta^2}e^{2i\beta\alpha},\\
&F_{\b,-\b,\b,-\b}=(\sin(\pi x))^{-2\beta^2},\\
&F_{\b,-\b,-\b,\b}=(2\tan(\frac{\pi x}{2}))^{-2\beta^2}.
\end{split}
\ee
The additional three terms in eq.~(\ref{G2psi}) can easily obtained from the results above with the replacement $\b\to{-\b}$,
\be
\begin{split}
&F_{-\b,-\b,\b,\b}=(2\cot{(\frac{\pi x}{2})})^{-2\beta^2}e^{-2i\beta\alpha},\\
&F_{-\b,\b,-\b,\b}=(\sin(\pi x))^{-2\beta^2},\\
&F_{-\b,\b,\b,-\b}=(2\tan(\frac{\pi x}{2}))^{-2\beta^2}.
\end{split}
\ee
Eventually, adding all the terms together, we obtain $G^{\Psi}_2(\boldsymbol{\alpha},x)$ 
	\be
	\begin{split}
	G^{\Psi}_2(\boldsymbol{\alpha},x)=a_1(e^{2i\beta\alpha}+e^{-2i\beta\alpha})+b_1
\end{split}
	\ee
 with $a_1=(2\cot{(\frac{\pi x}{2})})^{-2\beta^2}$ and $b_1=2(\sin{\pi x})^{-2\beta^2}+2(2\tan{\frac{\pi x}{2}})^{-2\beta^2}$. 
	Then $g_2^{\Psi}(\boldsymbol{\alpha},x)$ is given
	\be
	\begin{split}
	g_2^{\Psi}(\boldsymbol{\alpha},x)=\frac{2a_1\cos{(2\b\a})+b_1}{2a_1+b_1}.
\end{split}
	\ee
Here we have used the fact that $g_2^{I}(\bm{\a},x)=1$ since $[\rho_I,Q_A]=0$.
\par After the Fourier transform, we obtain the final result 
	\be
g^{\Psi}_2(x)=\frac{a_1\sin^2{(2\pi\b)}+2b_1\pi^2\b^2}{2\pi^2\b^2(2a_1+b_1)}.
	\ee
\section{Correlation functions of vertex and derivative operators }\label{appenB}
The strategy to compute the correlation functions involved derivative operators is to use the following trick to represent the current operator as a vertex operator
\be\label{dtr}
(i\partial \phi)(z)=\frac{1}{\e}\partial_z V_\e(z)|_{\e=0}
\ee 
\par Then terms appearing in eq.~(\ref{G2phi}) can be computed easily. For example,
\be
\begin{split}
F_{\b,-\b,J,J^{\dg}}&=\frac{\partial _{w_2^-,w_2^+}}{\e_1\e_2}\frac{\langle\mV_{\a}(i\infty)\mV_{-\a}(-i\infty)V_{\b}(w_1^-) V_{-\b}(w_1^+)\mV_{-\a}(i\infty)\mV_{\a}(-i\infty)V_{\e_1}(w_2^-) V^{\dg}_{\e_2}(w_2^+)\rangle_{\rm_{cyl}}}{\langle \mV_{\a}(i\infty)\mV_{-\a}(-i\infty)\mV_{-\a}(i\infty)\mV_{\a}(-i\infty)\rangle_{\rm_{cyl}}}\Big|_{\e_1,\e_2=0}\\
&=2^{-2-\b^2}(\csc(\frac{\pi x}{2}))^{\b^2}(\csc^2(\frac{\pi x}{2})+\b^2\tan^2(\frac{\pi x}{2})).
\end{split}
\ee
Similarly, one can get
\be
\begin{split}
&F_{\b,J^{\dg},J,-\b}=2^{-2-\b^2}(-\sec(\frac{\pi x}{2}))^{\b^2}(\b^2\cot^2(\frac{\pi x}{2})+\sec^2(\frac{\pi x}{2}))e^{i\a\b},\\
&F_{J,-\b,\b,J^{\dg}}=2^{-2-\b^2}(-\sec(\frac{\pi x}{2}))^{\b^2}(\b^2\cot^2(\frac{\pi x}{2})+\sec^2(\frac{\pi x}{2}))e^{-i\a\b},\\
&F_{J,J^{\dg},\b,-\b}=2^{-3-\b^2}(2-3\b^2+\b^2(\cos(\pi x)+2\sec^2(\frac{\pi x}{2}))(\sin(\frac{\pi x}{2}))^{-2-\b^2},\\
&F_{J,J^{\dg},J,J^{\dg}}=2^{-6}(7+\cos(2\pi x))^2\csc^4(\pi x).
\end{split}	
\ee
In  appendix \ref{appenA}, we have already obtained the value of $F_{\b,-\b,\b,-\b}=(\csc(\pi x))^{2\b^2}$. Adding all these terms together, we find
\be
\begin{split}
G_2(\boldsymbol{\alpha},x)&=a_2(e^{i\a\b}+e^{-i\a\b})+b_2.
\end{split}
\ee
where the coefficients $a_2$ and $b_2$ are given in eq.~(\ref{a2}) and eq.~(\ref{b2}) respectively.
Finally, one find that $g^{\Phi}_2(\a,x)$ has the same structure with $g^{\Psi}_2(\a,x)$
\be
g^{\Phi}_2(\a,x)=\frac{2a_2\cos{(\b\a})+b_2}{2a_2+b_2}.
\ee
After the Fourier transform,  the final result is obtained
\be
g^{\Phi}_2(x)=\frac{2a_2\sin^2{(\pi\b)}+b_2\pi^2\b^2}{\pi^2\b^2(2a_2+b_2)}
\ee
\section{RDMs and Correlation matrices in the XX spin chain}\label{appenC}
As is well known, the XX spin chain is described by the following Hamiltonian\cite{2001Book}
	\be
	H=-\frac{1}{4}\sum_{j=1}^L(\s^x_j\s^x_{j+1}+\s^y_j\s^y_{j+1}-h\s^z_j),
	\ee
	where $\s^{x,y,z}_j$ are the Pauli matrices acting on the $j$-th site.
	\par After a Jordan-Wigner transformation
	\be
	c_j=\left(\prod_{k=1}^{j-1}\s_k^z\right)\frac{\s_j^x-i\s_j^y}{2}\qquad c_j^{\dg}=\left(\prod_{k=1}^{j-1}\s_k^z\right)\frac{\s_j^x+i\s_j^y}{2},
	\ee
	this spin chain Hamiltonian is mapped to a free fermion Hamiltonian on the lattice
	\be
	H=-\frac12\sum_{j=1}^L\left[c_j^{\dg}c_{j+1}+c_{j+1}^{\dg}c_j-2h(c_j^{\dg}c_j-\frac12)\right],
	\ee
	where $c^{\dg}_j,c_j$ are fermionic creation and annihilation operators, satisfying the anticommutation relations $\{c_i,c^{\dg}_j\}=\d_{ij}$. Impose anti-periodic boundary conditions to the fermions, $c_{L+1}^{\dg}=-c_1^{\dg}, c_{L+1}=-c_1$. For simplicity, we will assume that $h=0$. 
	\par After Fourier transformation
	\be\label{cfb}
	b_k=\frac{1}{\sqrt{L}}\sum_{l=1}^Lc_le^{i\phi_k l},\qquad \phi_k\equiv\frac{2\pi k}{L}
	\ee
	the Hamiltonian can be diagonalized as\cite{2011Quantum} 
	\be
	H=\sum_{k\in\O}\e_k(b_k^{\dg}b_k-\frac12),
	\ee
	where $\e_k=-\cos k$, and $\O=\{\pm\frac{1}{2},\pm\frac{3}{2},\cdots\pm\frac{L-1}{2}\}$. The eigenstates of the Hamilton can be described by a set of momenta $K$,
	\be
	\ket{K}=\prod_{k\in K}b_k^{\dg}\ket{0}.
	\ee
	The ground state is half-filling with fermion number $n_F=\frac{L}{2}$ and is a Fermi sea with Fermi momentum $k_F=\frac{\pi}{2}$. And it's characterized by the set of momenta: $\{\pm\frac{1}{2},\pm\frac{3}{2},\cdots\pm\frac{n_F-1}{2}\}$. By removing or adding particles in momentum space close to the Fermi surface, we can obtain low-lying excited states\cite{Berganza:2011mh}. This model has a $U(1)$ symmetry generated by the conserved charge $Q=\sum^{L}_{j=1}c^{\dg}_jc_j$. 
	\par If the subsystem $A$ is made of $l$ contiguous lattice sites, then the RDM of the state $\ket{K}\bra{K}$ is given by\cite{2009Reduced} 
	\be
	\rho_A=\det{C_A^K}\exp\left(\sum_{i,j}\log[((C_A^K)^{-1}-1)]_{ij}c_i^{\dg}c_j\right).
	\ee
	where the $l\times l$ matrix $[C_A^K]_{m,n}=\bra{K}c_m^{\dg}c_n\ket{K}$, with $m,n\in A$, is the correlation matrix restricted in $A$. The elements of $C_A^K$ are given by 
	\be
	[C_A^K]_{mn}=\frac{1}{L}\sum_{k\in K}e^{i\phi_k(m-n)}.
	\ee
	\par It's convenient to introduce the Majorana fermionic operators\cite{Vidal:2002rm}
	\be\label{ac}
	a_{2s}=c^{\dg}_s+c_s,\qquad a_{2s-1}=i(c_s-c_s^{\dg}),
	\ee
	satisfying $\{a_i,a_j\}=2\d_{ij}$. For the case of a single interval with $l$ sites of the spin chain in a state $\ket{K}$, the Majorana correlation matrix can obtain
	\be
	\langle a_ia_j\rangle_K=\d_{ij}+\G^K_{ij},
	\ee
	with $\G\in\mc{M}_{2l}(\mathbb{C})$:
	\be\label{Gamma}
	\G^K=\begin{pmatrix}
		T^K_0&T^K_1&\cdots&T^K_{l-1}\\
		T^K_{-1}&T^K_0&\cdots&T^K_{l-2}\\
		\vdots&\vdots&\ddots&\vdots\\
		T^K_{1-l}&T^K_{2-l}&\cdots&T^K_0
	\end{pmatrix},
	\qquad T^K_{n}=\begin{pmatrix}
		f^K_{n}&g^K_{n}\\-g^K_{n}&f^K_{n}
	\end{pmatrix},
	\ee
	The correlation matrix $\G^K$ determines the $2^l\times2^l$ RDM. Moreover, we can also regard $e^{i\a Q_A}$ as some RDM with Majorana correlation matrix $\G^{\a}$ since
	\be
	\begin{split}
		&e^{i\a Q_A}=e^{i\a\sum_{j=1}^lc_j^{\dg}c_j}=\prod_{j=1}^l(e^{i\a}c_j^{\dg}c_j+c_jc_j^{\dg})\\
		&=(1+e^{i\a})^l\prod_{j=1}^l[p_jc_jc_j^{\dg}+(1-p_j)c_j^{\dg}c_j],
	\end{split}
	\ee
	where $p_j=\frac{1}{1+e^{i\a}}$. The Majorana correlation matrix $\G^\a$ and $\G^K$ (cf.~(\ref{Gamma})) have the same structure, but different block matrices\cite{Chen:2021pls}
	\be
	f^{\a}_{n}=0,\quad g^{\a}_{n}=\frac{i(1-e^{i\a})}{e^{i\a}+1}\d_{n,0}.
	\ee
	\par In the process of calculation, we will encounter the composition matrices indicated by $\G_i\times\G_j$. It is implicitly defined by\cite{2010Entanglement,Berganza:2011mh,1969Nonunitary} 
	\be
	\rho_{\G_i}\rho_{\G_j}=\tr(\rho_{\G_i}\rho_{\G_j})\rho_{\G_i\times\G_j},
	\ee
	where
	\be
	\tr(\rho_{\G_i}\rho_{\G_j})=\sqrt{\left|\frac{1+\G_i\G_j}{2}\right|},
	\ee
	and the product rule is
	\be
	\G_i\times\G_j=1-(1-\G_j)(1+\G_i\G_j)^{-1}(1-\G_i).
	\ee
	By associativity, the trace of the product of arbitrary number of RDMs can be obtained
	\be
	\tr(\r_{\G_i}\r_{\G_j}\cdots)=\tr(\r_{\G_i}\r_{\G_j})\tr(\r_{\G_i\times\G_j}\cdots).
	\ee
Based on the formula established above, one can numerically compute the charged moments $Z_n(\bm{\a})$ eq.~(\ref{Za}) with Hermitian $\rho_A$ for arbitrary $n$. Since this method is thoroughly studied in previous literature \cite{Chen:2021pls}, we will not discuss it here.  
\section{Non-Hermite fake RDMs and Correlation matrices}\label{appenD}
Considering the characteristics of the Majorana fermionic operators, we have
\be\label{aa4}
\begin{split}
	a_ia_j&=\begin{pmatrix}
	a_{2r}\\a_{2r-1}\end{pmatrix}\begin{pmatrix}
	a_{2s}a_{2s-1}\end{pmatrix}\\&=\begin{pmatrix} (c_r^{\dg}+c_r)(c_s^{\dg}+c_s)& i(c_r^{\dg}+c_r)(c_s^{\dg}-c_s)\\
	i(c_r^{\dg}-c_r)(c_s^{\dg}+c_s)& -(c_r^{\dg}-c_r)(c_s^{\dg}-c_s)\end{pmatrix}.
\end{split}
\ee
\par For excited state $\ket{\Psi}=\ket{V_1}+\ket{V_{-1}}$, the $S$ operator we have chosen is $S=-ia_{2n-1}a_{2n+1}$. Then the corresponding Majorana matrix can be computed as
\be
	\begin{split}
		f_1(r,s)={f_0(n,n+1)}^{-1}&(\langle V_{1}|c_n^{\dg}c^{\dg}_{n+1}c^{\dg}_rc_s|V_{-1}\rangle+\langle V_{1}|c_n^{\dg}c^{\dg}_{n+1}c^{\dg}_sc_r|V_{-1}\rangle\\&-\langle V_{1}|c_n^{\dg}c^{\dg}_rc^{\dg}_sc_{n+1}|V_{-1}\rangle-\langle V_{1}|c_{n+1}^{\dg}c^{\dg}_{r}c^{\dg}_sc_n|V_{-1}\rangle)
	\end{split}
\ee
	\be
	\begin{split}
		f_2(r,s)={f_0(n,n+1)}^{-1}&(\langle V_{1}|c_n^{\dg}c^{\dg}_{n+1}c^{\dg}_rc_s|V_{-1}\rangle+\langle V_{1}|c_n^{\dg}c^{\dg}_{n+1}c^{\dg}_sc_r|V_{-1}\rangle\\&+\langle V_{1}|c_n^{\dg}c^{\dg}_rc^{\dg}_sc_{n+1}|V_{-1}\rangle+\langle V_{1}|c_{n+1}^{\dg}c^{\dg}_{r}c^{\dg}_sc_n|V_{-1}\rangle)
	\end{split}
	\ee
	\be
	\begin{split}
		g_1(r,s)=-i{f_0(n,n+1)}^{-1}&(\langle V_{1}|c_n^{\dg}c^{\dg}_{n+1}c^{\dg}_rc_s|V_{-1}\rangle-\langle V_{1}|c_n^{\dg}c^{\dg}_{n+1}c^{\dg}_sc_r|V_{-1}\rangle\\&+\langle V_{1}|c_n^{\dg}c^{\dg}_rc^{\dg}_sc_{n+1}|V_{-1}\rangle+\langle V_{1}|c_{n+1}^{\dg}c^{\dg}_{r}c^{\dg}_sc_n|V_{-1}\rangle)+\d_{mn}
	\end{split}
	\ee
	\be
	\begin{split}
		g_2(r,s)=-i{f_0(n,n+1)}^{-1}&(-\langle V_{1}|c_n^{\dg}c^{\dg}_{n+1}c^{\dg}_rc_s|V_{-1}\rangle+\langle V_{1}|c_n^{\dg}c^{\dg}_{n+1}c^{\dg}_sc_r|V_{-1}\rangle\\&+\langle V_{1}|c_n^{\dg}c^{\dg}_rc^{\dg}_sc_{n+1}|V_{-1}\rangle+\langle V_{1}|c_{n+1}^{\dg}c^{\dg}_{r}c^{\dg}_sc_n|V_{-1}\rangle)+\d_{mn}
	\end{split}
	\ee
\par Let's first consider  $\langle V_{1}|c_n^{\dg}c^{\dg}_{n+1}c^{\dg}_rc_s|V_{-1}\rangle$. We have 
	\be
	\langle V_{1}|c_n^{\dg}c^{\dg}_{n+1}c^{\dg}_rc_s|V_{-1}\rangle=\frac{1}{L^2}\sum_{k,k',k_1,k_2\in\O}e^{-i\phi_kn-i\phi_{k'}(n+1)-i\phi_{k_1}r+i\phi_{k_2}s}\langle b^{\dg}_k b^{\dg}_{k'}b^{\dg}_{k_1}b_{k_2}b_qb_{q'}\rangle.
	\ee
Here and in the following, $\langle \cdot\rangle$ means the average under the state $V_1$. According to the Wick theorem, we have
	\be
	\begin{split}
		\langle b^{\dg}_k b^{\dg}_{k'}b^{\dg}_{k_1}b_{k_2}b_qb_{q'}\rangle&=\langle b_{k_1}^{\dg}b_{k_2}\rangle(\langle b_{k'}^{\dg}b_q\rangle\langle b_k^{\dg}b_{q'}\rangle-\langle b_{k}^{\dg}b_q\rangle\langle b_{k'}^{\dg}b_{q'}\rangle\\&-\langle b_{k'}^{\dg}b_{k_2}\rangle(\langle b_{k}^{\dg}b_q\rangle\langle b_{k_1}^{\dg}b_{q'}\rangle-\langle b_{k_1}^{\dg}b_q\rangle\langle b_{k}^{\dg}b_{q'}\rangle\\&+\langle b_{k}^{\dg}b_{k_2}\rangle(\langle b_{k'}^{\dg}b_q\rangle\langle b_{k_1}^{\dg}b_{q'}\rangle-\langle b_{k_1}^{\dg}b_q\rangle\langle b_{k'}^{\dg}b_{q'}\rangle\end{split}
	\ee
Using $\langle b_{k}^{\dg}b_{k'}\rangle=\delta_{k,k'}\d_{k\in V_1}$, we obtain
	\be
	\langle V_{1}|c_n^{\dg}c^{\dg}_{n+1}c^{\dg}_rc_s|V_{-1}\rangle=\frac{1}{L}\sum_{k\in V_1}[e^{-i\phi_{k}(r-s)}f_0(n,n+1)-e^{-i\phi_{k}(n+1-s)}f_0(n,r)+e^{-i\phi_{k}(n-s)}f_0(n+1,r)].
	\ee
The other terms can be obtained similarly
\be
\begin{split}
&\langle V_{1}|c_n^{\dg}c^{\dg}_{n+1}c^{\dg}_sc_r|V_{-1}\rangle=-\frac{1}{L}\sum_{k\in V_1}[e^{-i\phi_{k}(s-r)}f_0(n,n+1)-e^{-i\phi_{k}(n+1-r)}f_0(n,s)+e^{-i\phi_{k}(n-r)}f_0(n+1,s)],\\
&\langle V_{1}|c_n^{\dg}c^{\dg}_{r}c^{\dg}_sc_{n+1}|V_{-1}\rangle=\frac{1}{L}\sum_{k\in V_1}[e^{-i\phi_{k}(s-(n+1))}f_0(n,r)-e^{-i\phi_{k}(r-(n+1))}f_0(n,s)+e^{i\phi_{k}}f_0(r,s)],\\	
&\langle V_{1}|c_n^{\dg}c^{\dg}_{r}c^{\dg}_sc_{n+1}|V_{-1}\rangle=-\frac{1}{L}\sum_{k\in V_1}[e^{-i\phi_{k}(s-n)}f_0(n+1,r)-e^{-i\phi_{k}(r-n)}f_0(n+1,s)+e^{-i\phi_{k}}f_0(r,s)].
\end{split}
\ee
\par For excited state $\ket{\Phi}=\ket{V_1}+\ket{J}$, we should choose the $S$ operator as $S=a_{2n-1}$ instead. The corresponding Majorana matrix can be computed as
\be
\begin{split}
&\tilde{f}_1(r,s)={g_0(n)}^{-1}(-\langle V_1|c^{\dg}_rc^{\dg}_sc_n|J\rangle+\langle V_1|c^{\dg}_nc^{\dg}_rc_s|J\rangle+\langle V_1|c^{\dg}_nc^{\dg}_sc_r|J\rangle),\\
&\tilde{f}_2(r,s)={g_0(n)}^{-1}(\langle V_1|c^{\dg}_rc^{\dg}_sc_n|J\rangle+\langle V_1|c^{\dg}_nc^{\dg}_rc_s|J\rangle+\langle V_1|c^{\dg}_nc^{\dg}_sc_r|J\rangle),\\
&\tilde{g}_1(r,s)=-i{g_0(n)}^{-1}(\langle V_1|c^{\dg}_rc^{\dg}_sc_n|J\rangle+\langle V_1|c^{\dg}_nc^{\dg}_rc_s|J\rangle-\langle V_1|c^{\dg}_nc^{\dg}_sc_r|J\rangle)+\d_{rs},\\
&\tilde{g}_2(r,s)=-i{g_0(n)}^{-1}(\langle V_1|c^{\dg}_rc^{\dg}_sc_n|J\rangle-\langle V_1|c^{\dg}_nc^{\dg}_rc_s|J\rangle+\langle V_1|c^{\dg}_nc^{\dg}_sc_r|J\rangle)+\d_{rs}.
\end{split}
\ee
Now let's compute $\langle V_1|c^{\dg}_rc^{\dg}_sc_n|J\rangle$ first
\be
\langle V_1|c^{\dg}_rc^{\dg}_sc_n|J\rangle=\frac{i}{L^{\frac{3}{2}}}\sum_{k,k_1,k_2\in\O} e^{-i\phi_{k_1}r-i\phi_{k_2}s+i\phi_kn}\langle b_{k_1}^{\dg}b_{k_2}^{\dg}b_kb_q\rangle.
\ee
Applying the Wick's theorem, we get
\be
\begin{split}
	\langle V_1|c^{\dg}_rc^{\dg}_sc_n|J\rangle&=\frac{i}{L^{\frac{3}{2}}}\sum_{k,k_1,k_2\in\O} e^{-i\phi_{k_1}r-i\phi_{k_2}s+i\phi_kn}(\langle b_{k_2}^{\dg}b_q\rangle\langle b_{k_1}^{\dg}b_k\rangle-\langle b_{k_2}^{\dg}b_q\rangle\langle b_{k_1}^{\dg}b_k\rangle)\\&=\frac{i}{L^{\frac{3}{2}}}\sum_{k\in V_1} (e^{-i\phi_qs-i\phi_k(r-n)}-e^{-i\phi_qr-i\phi_k(s-n)})\\&=\frac{i}{L}\sum_{k\in V_1} (e^{-i\phi_k(r-n)}g_0(s)-e^{-i\phi_k(s-n)}g_0(r)).
\end{split}
\ee
The other terms can be obtained in the same way
\be
\langle V_1|c^{\dg}_nc^{\dg}_rc_s|J\rangle=\frac{i}{L}\sum_{k\in V_1} (e^{-i\phi_{k}(n-s)}g_0(r)-e^{-i\phi_{k}(r-s)}g_0(n)).
\ee
\be
\langle V_1|c^{\dg}_nc^{\dg}_sc_r|J\rangle=\frac{i}{L}\sum_{k\in V_1} (e^{-i\phi_{k}(s-r)}g_0(n)-e^{-i\phi_{k}(n-r)}g_0(s)).
\ee
\end{appendix}
\bibliographystyle{ieeetr}
\bibliography{Reference.bib}
\end{document}